\documentclass[11pt]{revtex4} 
\usepackage{amssymb,epsf}
\usepackage{latexsym}

\begin{document}

\title{Thermodynamical description of the interacting new agegraphic dark energy}
\author{A. Sheykhi $^{1,2}$\footnote{sheykhi@mail.uk.ac.ir} and M.R. Setare $^{3,2}$\footnote{rezakord@ipm.ir}}
\address{$^1$Department of Physics, Shahid Bahonar University, Kerman 76175, Iran\\
$^2$Research Institute for Astronomy and Astrophysics of Maragha
(RIAAM), Maragha,
         Iran\\
         $^3$  Department of Science, Payame Noor University, Bijar,
         Iran}
\begin{abstract}
We describe the thermodynamical interpretation of the interaction
between new agegraphic dark energy and dark matter in a non-flat
universe. When new agegraphic dark energy and dark matter evolve
separately, each of them remains in thermodynamic equilibrium. As
soon as an interaction between them is taken into account, their
thermodynamical interpretation changes by a stable thermal
fluctuation. We obtain a relation between the interaction term of
the dark components and this thermal fluctuation.
\end{abstract}
\maketitle

\section{Introduction\label{Int}}
The dark energy puzzle is one of the biggest challenges of the
modern cosmology in the past decade. There is an ample evidences
on the observational side that our universe is currently
experiencing a phase of accelerated expansion
\cite{Rie1,Rie2,Rie3,Rie4}. These observations suggest that nearly
three quarters of our universe consists of a mysterious energy
component (dark energy) which is responsible for this expansion,
and the remaining part consists of pressureless dark matter.
Nevertheless, despite the mounting observational evidences, the
nature of such dark energy remains elusive and it has become a
source of much debate except for the fact that it has negative
pressure. Most discussions on dark energy rely on the assumption
that it evolves independently of dark matter. Given the unknown
nature of both dark energy and dark matter there is nothing in
principle against their mutual interaction and it seems very
special that these two major components in the universe are
entirely independent. Indeed, this possibility has received a lot
of attention recently (see
\cite{Ame1,Ame2,Ame3,Ame4,Ame5,Zim1,Zim2,Zim3,Seta1,Set2,Set3,Set4,Set5,wang1,wang11,Shey0}
and references therein). In particular, it has been shown that the
coupling can alleviate the coincidence problem \cite{Pav1}.
Furthermore, it was argued that the appropriate coupling between
dark components can influence the perturbation dynamics and the
cosmic microwave background (CMB) spectrum and account for the
observed CMB low $l$ suppression \cite{wang2}. It was shown that
in a model with interaction the structure formation has a
different fate as compared with the non-interacting case
\cite{wang2}. It was also discussed that with strong coupling
between dark energy and dark matter, the matter density
perturbation is stronger during the universe evolution till today,
which shows that the interaction between dark energy and dark
matter enhances the clustering of dark matter perturbation
compared to the noninteracting case in the past. Therefore, the
coupling between dark components could be a major issue to be
confronted in studying the physics of dark energy. However, so
long as the nature of these two components remain unknown it will
not be possible to derive the precise form of the interaction from
first principles. Therefore, one has to assume a specific coupling
from the outset \cite{Ads,Amen1,Amen2} or determine it from
phenomenological requirements \cite{Chim1,Chim2}. Thermodynamical
description of the interaction (coupling) between holographic dark
energy and dark matter has been studied in \cite{wang3,SetVag}.

Among the various candidates to explain the accelerated expansion,
the agegraphic and new agegraphic dark energy (NADE) models
condensate in a class of quantum gravity may have interesting
cosmological consequences. These models take into account the
Heisenberg uncertainty relation of quantum mechanics together with
the gravitational effect in general relativity. The agegraphic
dark energy models assume that the observed dark energy comes from
the spacetime and matter field fluctuations in the universe
\cite{Cai1,Wei2,Wei1}. Since in agegraphic dark energy model the
age of the universe is chosen as the length measure, instead of
the horizon distance, the causality problem in the holographic
dark energy is avoided. The agegraphic models of dark energy  have
been examined and constrained by various astronomical observations
\cite{age1,age2,age3,age4,age5,age6,age7,sheykhi1,sheykhi2,sheykhi3,sheykhi4,sheykhi5,Setare2,Setare22}.
Although going along a fundamental theory such as quantum gravity
may provide a hopeful way towards understanding the nature of dark
energy, it is hard to believe that the physical foundation of
agegraphic dark energy is convincing enough. Indeed, it is fair to
say that almost all dynamical dark energy models are settled at
the phenomenological level, neither holographic dark energy model
nor agegraphic dark energy model is exception. Though, under such
circumstances, the models of holographic and agegraphic dark
energy, to some extent, still have some advantage comparing to
other dynamical dark energy models because at least they originate
from some fundamental principles in quantum gravity.

The main purpose of this Letter is to study thermodynamical
interpretation of the interaction between dark matter and NADE
model for a universe enveloped by the apparent horizon. It was
shown that for an accelerating universe the apparent horizon is a
physical boundary from the thermodynamical point of view
\cite{Jia}. In particular, it was argued that for an accelerating
universe inside the event horizon the generalized second law does
not satisfy, while the accelerating universe enveloped by the
apparent horizon satisfies the generalized second law  of
thermodynamics \cite{Jia,sheywang1,sheywang2,sheywang3}.
Therefore, the event horizon in an accelerating universe might not
be a physical boundary from the thermodynamical point of view.
This Letter is outlined as follows. In the next section we
consider the thermodynamical picture of the non-interacting NADE
in a non-flat universe. In section \ref{Int}, we extend the
thermodynamical description in the case where there is an
interaction term between the dark components. We also present an
expression for the interaction term in terms of a thermal
fluctuation. The last section is devoted to summary and
discussion.

\section{Thermodynamical description of the non-interacting NADE  \label{NonInt}}
We consider the Friedmann-Robertson-Walker (FRW) universe which is
described by the line element
\begin{eqnarray}
 ds^2=dt^2-a^2(t)\left(\frac{dr^2}{1-kr^2}+r^2d\Omega^2\right),\label{metric}
 \end{eqnarray}
where $a(t)$ is the scale factor, and $k$ is the curvature
parameter with $k = -1, 0, 1$ corresponding to open, flat, and
closed universes, respectively. A closed universe with a small
positive curvature ($\Omega_k\simeq0.01$) is compatible with
observations \cite{spe1,spe2,spe3,spe4}. The first Friedmann
equation takes the form
\begin{eqnarray}\label{Fried}
H^2+\frac{k}{a^2}=\frac{1}{3m_p^2} \left( \rho_m+\rho_D \right),
\end{eqnarray}
where $H=\dot{a}/a$ is the Hubble parameter, $\rho_{m}$ and
$\rho_{D}$ are the energy density of dark matter and dark energy,
respectively. We define, as usual, the fractional energy densities
such as
\begin{eqnarray}\label{Omega}
\Omega_m=\frac{\rho_m}{3m_p^2H^2}, \hspace{0.5cm}
\Omega_D=\frac{\rho_D}{3m_p^2H^2},\hspace{0.5cm}
\Omega_k=\frac{k}{H^2 a^2}.
\end{eqnarray}
Thus, the Friedmann equation can be written
\begin{eqnarray}\label{Fried2}
\Omega_m+\Omega_D=1+\Omega_k.
\end{eqnarray}
Let us first review the origin of the agegraphic dark energy
model. Following the line of quantum fluctuations of spacetime,
Karolyhazy et al. \cite{Kar1,Kar2,Kar3} argued that the distance
$t$ in Minkowski spacetime cannot be known to a better accuracy
than $\delta{t}=\beta t_{p}^{2/3}t^{1/3}$ where $\beta$ is a
dimensionless constant of order unity. Based on Karolyhazy
relation, Sasakura discussed that the energy density of metric
fluctuations of the Minkowski spacetime is given by \cite{Sas}
(see also \cite{Maz1,Maz2})
\begin{equation}\label{rho0}
\rho_{D} \sim \frac{1}{t_{p}^2 t^2} \sim \frac{m^2_p}{t^2},
\end{equation}
where $t_{p}$ is the reduced Planck time and $t$ is a proper time
scale. On these basis, Cai wrote down the energy density of the
original agegraphic dark energy as \cite{Cai1}
\begin{equation}\label{rho1}
\rho_{D}= \frac{3n^2 m_{p}^2}{T_A^2},
\end{equation}
where $T_A$ is the age of the universe,
\begin{equation}
T_A=\int_0^a{\frac{da}{Ha}},
\end{equation}
and the numerical factor $3n^2$ is introduced to parameterize some
uncertainties, such as the species of quantum fields in the
universe, the effect of curved space-time, and so on. However, to
avoid some internal inconsistencies in the original agegraphic
dark energy model, the so-called ``new agegraphic dark energy" was
proposed, where the time scale is chosen to be the conformal time
$\eta$ instead of the age of the universe \cite{Wei2}. The NADE
contains some new features different from the original agegraphic
dark energy and overcome some unsatisfactory points. For instance,
the original agegraphic dark energy suffers from the difficulty to
describe the matter-dominated epoch while the NADE resolved this
issue \cite{Wei2}. The energy density of the NADE can be written
\begin{equation}\label{rho1new}
\rho_{D}= \frac{3n^2 m_{p}^2}{\eta^2},
\end{equation}
where the conformal time $\eta$ is given by
\begin{equation}
\eta=\int{\frac{dt}{a}}=\int_0^a{\frac{da}{Ha^2}}.
\end{equation}
Consider the FRW universe filled with dark energy and dust (dark
matter) which evolve according to their conservation laws
\begin{eqnarray}
&&\dot{\rho}_D+3H\rho_D(1+w_D^0)=0,\label{consq1}\\
&&\dot{\rho}_m+3H\rho_m=0, \label{consm1}
\end{eqnarray}
where $w_D=p_D/\rho_D$ is the equation of state parameter of NADE.
The superscript above the equation of state parameter, $w_D$,
denotes that there is no interaction between the dark components.
The fractional energy density of the NADE is given by
\begin{eqnarray}\label{Omegaqnew}
\Omega_D=\frac{n^2}{H^2\eta^2},
\end{eqnarray}
where its evolution behavior is governed by \cite{sheykhi1}
\begin{eqnarray}\label{Omegaq3}
{\Omega'_D}&=&\Omega_D\left[(1-\Omega_D)\left(3-\frac{2}{n}\sqrt{\Omega_D}\right)
+\Omega_k\right].
\end{eqnarray}
Here the prime stands for the derivative with respect to
$x=\ln{a}$. Taking the  derivative with respect to the cosmic time
of Eq. (\ref{rho1new}) and using Eq. (\ref{Omegaqnew}) we get
\begin{eqnarray}\label{rhodotnew}
\dot{\rho}_D=-2H\frac{\sqrt{\Omega_D^0}}{na}\rho_D.
\end{eqnarray}
Inserting this relation into Eq. (\ref{consq1}) we obtain the
equation of state parameter of the NADE
\begin{eqnarray}\label{wqnew}
1+w_D^0=\frac{2}{3na}\sqrt{\Omega_D^0}.
\end{eqnarray}
We also limit ourselves to the assumption that the thermal system
bounded by the apparent horizon remains in equilibrium so that the
temperature of the system must be uniform and the same as the
temperature of its boundary. This requires that the temperature
$T$ of the energy content inside the apparent horizon should be in
equilibrium with the temperature $T_h$ associated with the
apparent horizon, so we have $T=T_h$. This expression holds in the
local equilibrium hypothesis. If the temperature of the fluid
differs much from that of the horizon, there will be spontaneous
heat flow between the horizon and the fluid and the local
equilibrium hypothesis will no longer hold. This is also at
variance with the FRW geometry. Thus, when we consider the thermal
equilibrium state of the universe, the temperature of the universe
is associated with the horizon temperature. In this picture the
equilibrium entropy of the NADE is connected with its energy and
pressure through the first  law of thermodynamics
\begin{equation}\label{Gib1}
T dS_D=dE_D+p_DdV,
\end{equation}
where the volume enveloped by the apparent horizon is given by
\begin{equation}\label{V}
V=\frac{4\pi}{3}r_A^3,
\end{equation}
and $r_A$ is the apparent horizon radius. The apparent horizon was
argued as a causal horizon for a dynamical spacetime and is
associated with gravitational entropy and surface gravity
\cite{Hay1,Hay2,Bak}. For the FRW universe the apparent horizon
radius reads \cite{sheyahmad1,sheyahmad2}
\begin{equation}
\label{radius}
 {r}_A=\frac{1}{\sqrt{H^2+k/a^2}}.
\end{equation}
The total energy of the NADE inside the apparent horizon is
\begin{equation}\label{E}
E_D=\rho_D V= \frac{4\pi n^2m_{p}^2r_A^3}{\eta^2}.
\end{equation}
Taking the differential form of Eq. (\ref{E}) and using Eq.
(\ref{Omegaqnew}), we find
\begin{equation}\label{E2}
dE_D=4\pi
m_{p}^2({{r}^0_A})^2H_0^2\Omega_D^0\left[3d{r}^0_A-2\frac{{r}^0_A}{n}H_0\sqrt{\Omega_D^0}d\eta^0
\right].
\end{equation}
The associated temperature on the apparent horizon can be written
as
\begin{equation}\label{Therm}
T =\frac{1}{2\pi r_A}.
\end{equation}
Inserting Eqs. (\ref{V}),  (\ref{E2}) and (\ref{Therm}) into
(\ref{Gib1}), we obtain
\begin{equation}\label{ds1}
dS_D^{(0)} =8\pi^2
m_{p}^2({{r}^0_A})^3H_0^2\Omega_D^0\left[3(1+w_D^0)d{r}^0_A-2\frac{{r}^0_A}{n}H_0\sqrt{\Omega_D^0}d\eta^0
\right],
\end{equation}
Using Eq. (\ref{wqnew}) as well as relation $H_0
d\eta^0=dx^0/a_0$, we find
\begin{equation}\label{ds2}
dS_D^{(0)} =16\pi^2 m_{p}^2({{r}^0_A})^3H_0^2\Omega_D^0
\frac{\sqrt{\Omega_D^0}}{na_0}\left[d{r}^0_A-{r}^0_A dx^0 \right].
\end{equation}
Here the superscript/subscript $(0)$ denotes that in this picture
our universe is in a thermodynamical stable equilibrium.
\section{Thermodynamical description of the interacting NADE \label{Int}}
In this section we study the case where the pressureless dark
matter and the NADE interact with each other. In this case
$\rho_{m}$ and $\rho_{D}$ do not conserve separately; they must
rather enter the energy balances
\begin{eqnarray}
&&\dot{\rho}_m+3H\rho_m=Q, \label{consm2}
\\&& \dot{\rho}_D+3H\rho_D(1+w_D)=-Q.\label{consq2}
\end{eqnarray}
Here $Q$  denotes the interaction term and can be taken as $Q
=3b^2 H(\rho_m+\rho_{D})$  with $b^2$  being a coupling constant
\cite{Pav1}. Inserting Eq. (\ref{rhodotnew}) into (\ref{consq2}),
we obtain the equation of state parameter of the interacting NADE
\begin{eqnarray}\label{wq2}
1+w_D=\frac{2}{3na}\sqrt{\Omega_D}-\frac{Q}{9m_{p}^2 H^3
\Omega_D}.
\end{eqnarray}
The evolution behavior of the NADE is now given by \cite{sheykhi1}
\begin{eqnarray}\label{Omegaq3new}
{\Omega'_D}&=&\Omega_D\left[(1-\Omega_D)\left(3-\frac{2}{na}\sqrt{\Omega_D}\right)
-3b^2(1+\Omega_k)+\Omega_k\right].
\end{eqnarray}
Comparing Eq. (\ref{wq2}) with Eq. (\ref{wqnew}), we see that the
presence of the interaction term $Q$ has provoked a change in the
equation of state parameter and consequently in the dimensionless
density parameter of the dark energy component and thus now there
is no subscript above the aforesaid quantities to denote the
absence of interaction. The interacting NADE model in the non-flat
universe as described above is not anymore thermodynamically
interpreted as a state in thermodynamical equilibrium. Indeed, as
soon as an interaction between dark components is taken into
account, they cannot remain in their respective equilibrium
states.  The effect of interaction between the dark components is
thermodynamically interpreted as a small fluctuation around the
thermal equilibrium. It was shown \cite{Das} that due to the
fluctuation, there is a leading logarithmic correction
$S^{(1)}_D=-\frac{1}{2}\ln(CT^2)$ to the thermodynamic entropy
around equilibrium in all thermodynamical systems. Therefore, the
entropy of the NADE is connected with its energy and pressure
through the first law of thermodynamics
\begin{equation}\label{Gib2}
T dS_D=dE_D+p_DdV,
\end{equation}
where now the entropy has been assigned an extra logarithmic
correction \cite{Das}
\begin{equation}\label{S}
S_D=S^{(0)}_D+S^{(1)}_D,
\end{equation}
where the leading logarithmic correction is
\begin{equation}\label{S1}
S^{(1)}_D=-\frac{1}{2}\ln(CT^2),
\end{equation}
and $C$ is the heat capacity defined by
\begin{equation}\label{C}
C=T\frac{\partial S^{(0)}_D}{\partial T}.
\end{equation}
It is a matter of calculation to show that
\begin{equation}\label{C1}
C=-16\pi^2 m_{p}^2({{r}^0_A})^4H_0^2\Omega_D^0
\frac{\sqrt{\Omega_D^0}}{na_0},
\end{equation}
and therefore
\begin{equation}\label{SS1}
S^{(1)}_D=-\frac{1}{2}\ln\left(-4
m_{p}^2({{r}^0_A})^2H_0^2\Omega_D^0
\frac{\sqrt{\Omega_D^0}}{na_0}\right).
\end{equation}
Substituting the expressions for the volume, energy, and
temperature in Eq. (\ref{Gib2}) for the interacting case, we get
\begin{equation}\label{dsint1}
dS_D =8\pi^2 m_{p}^2{{r}^3_A}
H^2\Omega_D\left[3(1+w_D)d{r}_A-\frac{2{r}_A}{n}H\sqrt{\Omega_D}d\eta
\right],
\end{equation}
or in another way
\begin{equation}\label{dsint2}
dS_D =8\pi^2 m_{p}^2{{r}^3_A}
H^2\Omega_D\left[3(1+w_D)d{r}_A-\frac{2{r}_A}{na}\sqrt{\Omega_D}dx
\right],
\end{equation}
and thus one gets
\begin{eqnarray}\label{dsint3}
1+w_D&=&\frac{1}{24\pi^2 m_{p}^2{{r}^3_A}
H^2\Omega_D}\frac{dS_D}{d{r}_A}+\frac{2{r}_A}{3na}\sqrt{\Omega_D}\frac{dx}{d{r}_A}, \nonumber \\
&&=\frac{1}{24\pi^2 m_{p}^2{{r}^3_A}
H^2\Omega_D}\left[\frac{dS^{(0)}_D}{d{r}_A}+\frac{dS^{(1)}_D}{d{r}_A}\right]+\frac{2{r}_A}{3na}\sqrt{\Omega_D}\frac{dx}{d{r}_A}.
\end{eqnarray}
Employing Eqs. (\ref{ds2}), (\ref{S1})-(\ref{SS1}), we can easily
find
\begin{eqnarray}\label{ds0}
\frac{dS^{(0)}_D}{d{r}_A}&=&\frac{\partial
S^{(0)}_D}{\partial{r}^0_A}\frac{d{r}^0_A}{d{r}_A}+\frac{\partial
S^{(0)}_D}{\partial{x}^0}\frac{d{x}^0}{d{r}_A}=16\pi^2
m_{p}^2{({r_A^0})^3} H_0^2\frac{({\Omega_D^0})^{3/2}}{na_0}
\left(\frac{d{r}^0_A}{d{r}_A}-{r}^0_A\frac{d{x}^0}{d{r}_A}\right),\\
\frac{dS^{(1)}_D}{d{r}_A}&=&\frac{\partial
S^{(1)}_D}{\partial{r}^0_A}\frac{d{r}^0_A}{d{r}_A}=-\frac{1}{{r}^0_A}\frac{d{r}^0_A}{d{r}_A}.\label{ds1}
\end{eqnarray}
Finally, by equating expressions (\ref{wq2}) and (\ref{dsint3})
for the equation of state parameter of the interacting NADE
evaluated on cosmological and thermodynamical sides, respectively,
one gets an expression for the interaction term
\begin{eqnarray}\label{QF}
\frac{Q}{9m_{p}^2 H^3}&=&\frac{2\sqrt{\Omega_D}}{3na}
{\Omega_D}\left(1-r_A\frac{dx}{dr_A}\right)-\frac{1}{24\pi^2
m_{p}^2{{r_A}^3}
H^2}\left[\frac{dS^{(0)}_D}{dr_A}+\frac{dS^{(1)}_D}{dr_A}\right]\nonumber
\\ &=&\frac{2\sqrt{\Omega_D}}{3na}
{\Omega_D}\left(1-r_A\frac{dx}{dr_A}\right)-\frac{2}{3na_0}\frac{H_0^2}{H^2}({\Omega_D^0})^{3/2}
\left(\frac{{r}^0_A}{{r}_A}\right)^3\left(\frac{d{r}^0_A}{d{r}_A}-{r}^0_A\frac{d{x}^0}{d{r}_A}\right)
\nonumber \\ &&+\frac{1}{24\pi^2 m_{p}^2{{r_A}^3}{r}^0_A
H^2}\frac{d{r}^0_A}{d{r}_A} .
\end{eqnarray}
In this way we provide the relation between the interaction term
of the dark components and the thermal fluctuation.
\section{Summary and discusion\label{CONC}}
 One of the important questions concerns the thermodynamical
behavior of the accelerated expanding universe driven by dark
energy. It is interesting to ask whether thermodynamics in an
accelerating universe can reveal some properties of dark energy.
It was first pointed out in \cite{Jac} that the hyperbolic second
order partial differential Einstein equation has a predisposition
to the first law of thermodynamics. The profound connection
between the thermodynamics and the gravitational field equations
has also been observed in the cosmological situations
\cite{Cai2,Cai3,CaiKim,Wang1,Wang2,Wang3,Cai4,sheyahmad3}. This
connection implies that the thermodynamical properties can help
understand the dark energy, which gives strong motivation to study
thermodynamics in the accelerating universe.

Although at this point the interaction between dark energy and
dark matter looks purely phenomenological, but in the absence of a
symmetry that forbids the interaction there is nothing, in
principle, against it. Further, the interacting dark mater–dark
energy (the latter in the form of a quintessence scalar field and
the former as fermions whose mass depends on the scalar field) has
been investigated at one quantum loop with the result that the
coupling leaves the dark energy potential stable if the former is
of exponential type but it renders it unstable otherwise. Thus,
microphysics seems to allow enough room for the coupling; however,
this point is not fully settled and should be further
investigated. Recently evidence was provided by the Abell Cluster
A$586$ in support of the interaction between dark energy and dark
matter \cite{Ber1,Ber2}.

In this Letter, we provided a thermodynamical description for the
NADE  model in a universe with spacial curvature. It was shown
that for an accelerating universe the apparent horizon is a
physical boundary from the thermodynamical point of view. We
explored the thermodynamical picture of the interacting NADE model
for a FRW universe enveloped by the apparent horizon. The NADE
contains some new features different from the original agegraphic
dark energy and overcome some unsatisfactory points. For instance,
the original agegraphic dark energy suffers from the difficulty to
describe the matter-dominated epoch while the NADE resolved this
issue. We assumed that in the absence of a coupling, the two dark
components remain in separate thermal equilibrium and that the
presence of a small coupling between them can be described as
stable fluctuations around equilibrium. Finally, resorting to the
logarithmic correction to the equilibrium entropy we derived an
expression for the interaction term in terms of a thermal
fluctuation.

\acknowledgments{This work has been supported by Research
Institute for Astronomy and Astrophysics of Maragha.}

\end{document}